\documentstyle[epsfig,longtable]{aipproc}

\def\gs{\mathrel{\raise0.35ex\hbox{$\scriptstyle >$}\kern-0.6em 
\lower0.40ex\hbox{{$\scriptstyle \sim$}}}}
\def\ls{\mathrel{\raise0.35ex\hbox{$\scriptstyle <$}\kern-0.6em 
\lower0.40ex\hbox{{$\scriptstyle \sim$}}}}
\begin{document}

\title{Deep sub-mm surveys with SCUBA}
 
\author{Ian Smail,$^{\! 1}$ Rob Ivison,$^{\! 2}$
Andrew Blain$^3$ \& Jean-Paul Kneib$^4$}
\address{$^1$Department of Physics, University of Durham,
South Road, Durham DH1 3LE\\
$^2$ Dept.\ of Physics \& Astronomy, University College London, 
London WC1E 6BT\\
$^3$ Cavendish Laboratory, Madingley Road, Cambridge CB3 0HE\\
$^4$ Observatoire de Toulouse, 14 Avenue E.\ Belin,
31400 Toulouse, France}

\lefthead{Deep SCUBA Surveys}
\righthead{Smail et al.}
\maketitle

\vspace*{-6mm}
\begin{abstract}
We review published deep surveys in the submillimeter (sub-mm) regime
from the new Sub-millimetre Common User Bolometer Array (SCUBA,
\cite{H99}) on the 15-m James Clerk Maxwell Telescope (JCMT), Mauna
Kea, Hawaii.  Summarising the number counts of faint sub-mm sources
determined from the different surveys we show that the deepest counts
from our completed SCUBA Lens Survey, down to 0.5\,mJy at 850\,$\mu$m,
fully account for the far-infrared background (FIRB) detected by {\it
COBE}.  We conclude that a population of distant, dust-enshrouded
ultraluminous infrared galaxies dominate the FIRB emission around
1\,mm.  We go on to discuss the nature of this population, starting
with the identification of their optical counterparts, where we
highlight the important role of deep VLA radio observations in this
process.  Taking advantage of the extensive archival {\it Hubble Space
Telescope} ({\it HST}\,) observations of our fields, we then
investigate the morphological nature of the sub-mm galaxy population
and show that a large fraction exhibit disturbed or interacting
morphologies.  By employing existing broadband photometry, we derive
crude redshift limits for a complete sample of faint sub-mm galaxies
indicating that the majority lie at $z<5$, with at most 20\% at higher
redshifts.  We compare these limits to the initial spectroscopic
results from various sub-mm samples.  Finally we discuss the nature of
the sub-mm population, its relationship to other classes of
high-redshift galaxies and its future role in our understanding of the
formation of massive galaxies.
\end{abstract}

\vspace*{-6mm}
\subsection*{Introduction}
\vspace*{-2mm}
The extragalactic background light is the repository for all emission
from the distant Universe and thus contains unique information about
the star-formation history of the Universe.  The far-infrared component
of this (the FIRB) was detected by {\it COBE} (e.g.\ \cite{P96}) at a
level comparable to that seen in the optical background\cite{RAB},
suggesting that a large proportion of the stars seen in the local
Universe were formed in dust-obscured galaxies at high
redshifts\cite{BSIK}.  Such strongly star-forming, dusty, distant
galaxies would be luminous sub-mm sources, due to the re-radiation in
the rest-frame sub-mm of the UV/optical starlight absorbed by the
dust.  Thus deep sub-mm observations, at $\lambda \gs 100\,\mu$m, would
be a fruitful avenue to pursue in the search for these forming
galaxies.  The strong negative $K$-corrections provided by the thermal
dust spectrum of galaxies also means these systems are easily
observable out to high redshift\cite{HD97}.  A {\it doppelg\"anger} for
Arp\,220, with an star-formation rate (SFR) of $\gs 100$\,M$_\odot$
yr$^{-1}$, would have a 850-$\mu$m flux density of $\gs
3$\,mJy\footnote{We assume $q_o=0.5$ and $h_{\rm 100}=0.5$ throughout.}
out to $z\sim 10$ ($\gs 0.3$\,mJy for $q_o=0.05$\cite{HD97}).  The
recently commissioned SCUBA camera on JCMT can achieve this flux limit
across a 5 sq.\ arcmin field in one night.

\subsection*{Faint Galaxy Counts in the Sub-mm}
\vspace*{-2mm}
At the time of writing, 850-$\mu$m counts of faint extragalactic
sources have been published by four groups\cite{SIB,Hu98,B98,Ea98}.
The first indications of the surface density of mJy 850-$\mu$m galaxies
was given by Smail, Ivison \& Blain\cite{SIB}, who took advantage of
gravitational amplification by massive cluster lenses to increase the
sensitivity of their SCUBA maps, and derived a source count of $(2.5\pm
1.4)\times 10^3$\,deg$^{-2}$ down to a flux density limit of 4\,mJy on
the basis of 6 detections.  This surface density has been broadly
confirmed by a number of subsequent studies of blank fields, which
spurn lens amplification, preferring simple brute-force integration to
obtain the necessary sensitivity.\footnote{Several other groups are
pursuing surveys of lensing clusters using SCUBA, including those
headed by Scott Chapman and Paul van der Werf, first results from these
should appear soon.}    These studies include maps of the Canada-France
Redshift Survey fields (CFRS, \cite{Ea98}), the Lockman Hole and Hawaii
Survey Fields\cite{B98} and the {\it Hubble Deep Field}\cite{Hu98},
and detect 11, 2 and 5 sources above their respective flux limits.
The latter two surveys reach the blank-field confusion limit of the
JCMT\cite{BIS} in their deepest integrations.  The sub-mm source
densities derived from the different surveys are plotted in Fig.~1 to
show the broad level of agreement reached.

The latest results from the analysis of the completed SCUBA Lens
Survey\cite{SIBK} are also shown in Fig.~1.  The complete sample
comprises a total of 17 galaxies detected at 3\,$\sigma$ significance
or above, and 10 detected above 4\,$\sigma$, in the fields of seven
massive and well-studied cluster lenses at $z=0.19$--$0.41$. The total
surveyed area is 0.01\,degree$^{2}$, with a sensitivity of better than
2\,mJy in the image planes.  The analysis of these
catalogues\cite{BKIS} makes use of well-constrained lens models for all
the clusters (e.g.\ \cite{JPK93}) to accurately correct the observed
source fluxes for lens amplification.  For the median source
amplification, $\sim 2.5\times$, our survey covers an area of the
source plane equivalent to roughly three times the SCUBA {\it HDF} map
at a comparable sensitivity and with a factor of two finer beam size.
At higher amplifications, the survey covers a smaller region, but at a
correspondingly higher sensitivity (e.g.\ $\sim 1$ sq.\ arcmin at
$\sigma_{850} \sim 0.1$\,mJy) and resolution.  The uncertainties
associated with our lensing analysis are included in the final error
quoted on the derived counts.  {\it The total uncertainty in the
lensing correction is at most comparable to the typical error in the
absolute SCUBA calibration.} The magnification produced by the massive
cluster lenses allows us to constrain the source counts down to
0.5\,mJy\cite{BKIS}, four times fainter than the deepest blank-field
counts published.  Moreover, these observations are less affected by
confusion noise, due to the expanded view of the background source
plane provided by the cluster lenses.

At 450\,$\mu$m, which SCUBA provides simultaneously with the 850-$\mu$m
maps, only a few sources have been detected in any of the published
surveys.  This is due to a combination of the lower atmospheric
transmission at 450\,$\mu$m in normal conditions on Mauna Kea, the
lower efficiency of the JCMT dish surface at 450\,$\mu$m and the
relatively high redshifts of the bulk of the sub-mm population\cite{Hu98}.

%
% Figure 1
%
\begin{figure}[htb]
\centerline{\epsfig{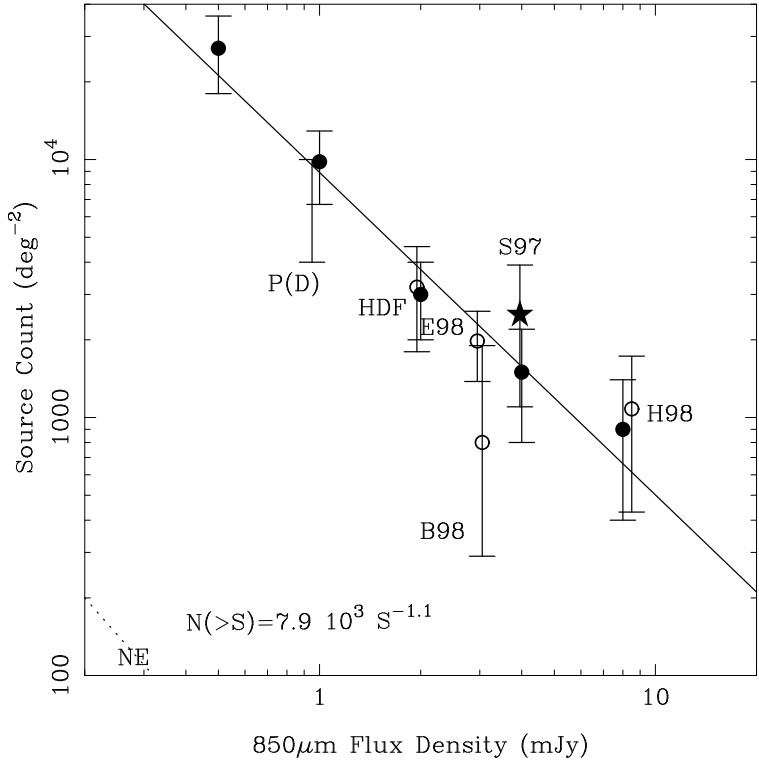}}
\vspace*{0.3cm}
\caption{
A comparison of 850-$\mu$m galaxy counts from different groups (S97
[6]; {\it HDF} and $P(D)$ [7]; B98 [8]; E98 [9]; H98 [11]).  The
filled circles come from the analysis of the completed SCUBA Lens Survey
[13] and are corrected for lens amplification.  The solid line shows
the fit to all of the observations with the form $N(>S) = 7.9 \times
10^3 S^{-1.1}$, which is an adequate description of the counts at flux
densities of 0.5--10\,mJy.  The dotted line (labelled NE) indicates the
count expected on the basis of a non-evolving local {\it IRAS} luminosity
function.
}
\end{figure}

The cumulative 850-$\mu$m counts of Smail, Ivison \& Blain (1997)
accounted for roughly 30\% of the FIRB detected by {\it COBE}
(e.g.\ \cite{P96,Fi98}).  The counts from the HDF brighter than the
confusion limit at 2\,mJy account for close to 50\% of the FIRB, while
the deepest counts from the lens fields\cite{BKIS} indicate that the
bulk of the FIRB is resolved at 0.5\,mJy.  This suggests that not only
must the counts converge around 0.5\,mJy, but that the FIRB is
dominated by emission from the most luminous sources.  This is a
remarkable achievement given that the FIRB was detected only three
years ago and SCUBA has been operating for a little over a year.

Having resolved the background we can now study the nature of the
populations contributing to the FIRB and so determine at what epoch the
background was emitted.  Here again our survey has the advantage of
lens amplification, this time in the optical and near-IR where the
identification and spectroscopic follow-up are undertaken.  Typically
the counterparts of our sub-mm sources will appear $\sim 1$\,magnitude
brighter than the equivalent galaxy in a blank field.

%
% Figure 2
%
\begin{figure}[htb]
\centerline{\epsfig{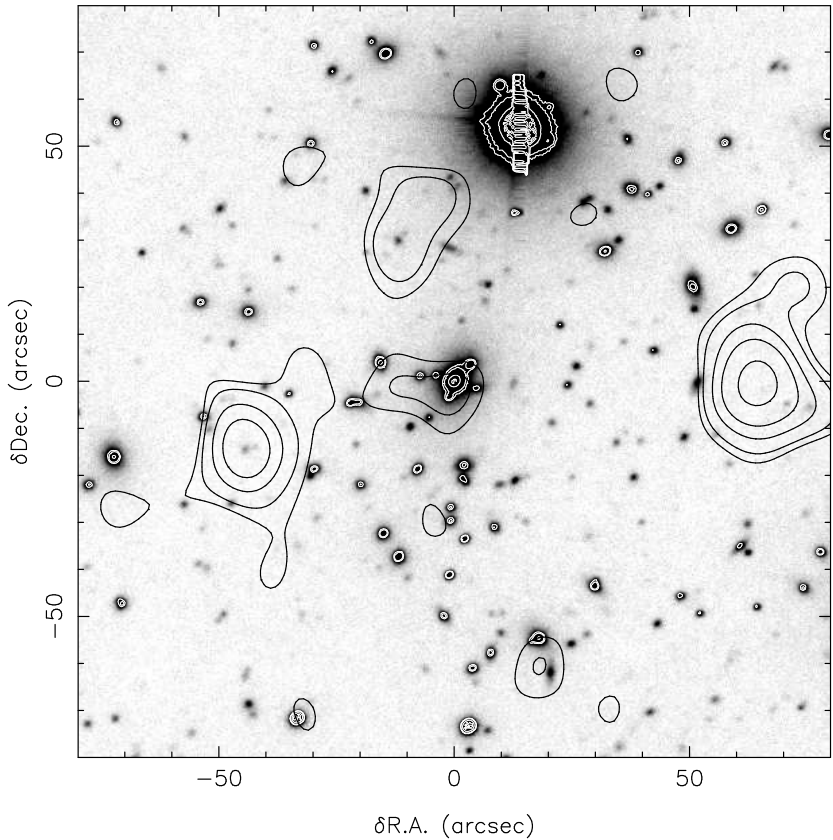}}
\vspace*{0.3cm}
\caption{
The SCUBA 850-$\mu$m map of the rich cluster A\,1835 ($z=0.25$), this
is overlayed on a deep $I$-band image of the cluster taken with the
Palomar 5.1-m Hale.  Two bright sub-mm sources are visible either side
of the central cluster galaxy, both are coincident with $>
100$-$\mu$Jy radio sources in our deep VLA map [17]. The eastern source is
identified with an $I\sim 21$ interacting galaxy at $z=2.6$ [16,17],
while the western source has no obvious counterpart at the position of
the sub-mm/radio peak.  Weak sub-mm emission is also detected from
vigorous star formation in the central galaxy of this cooling-flow
cluster [18]. 
}
\end{figure}

\vspace*{-6mm}
\subsection*{Identifications and Morphologies of SCUBA Galaxies}
\vspace*{-2mm}
The sub-mm fluxes of the sources detected in all the published surveys
are in the range $S_{850} \sim 0.5$--$10$\,mJy, equivalent to
luminosities of $\log_{10} L_{\rm FIR} \sim 12$--$13$ if they lie at
$z\gs 1$, and so they class as ultraluminous infrared galaxies
(ULIRGs).  Smail et al.\ (1998, \cite{SIBK}) presented optical
identifications obtained from deep {\it HST}\, and ground-based images
for galaxies selected from the SCUBA Lens survey (e.g.\ Fig.~2).  Down
to a limit of $I\sim 25$, counterparts are identified for 14 of the 16
sources in the 3-$\sigma$ sample and for 9/10 sources in the 4-$\sigma$
catalog that lie within the optical fields.  This rate of optical
identification, 80--90\%, down to $I\sim 25$ is similar to that
achieved by the other sub-mm surveys\cite{Hu98,Ea98}.   The bulk of
these sources are resolved in the optical images indicating that they
are galaxies

We are undertaking near-IR imaging of all our fields to search for any
extremely red counterparts which could have been missed in the optical
identifications.  A link has been suggested between the population of
extremely red objects (EROs) and the sub-mm galaxies due to the recent
sub-mm detection of one of the most well-studied EROs, HR10\cite{D98}.
We have so far identified only one possible ERO counterpart to a sub-mm
source in our survey with the bulk of the sub-mm galaxies showing
optical--near-IR colors more typical of the general field, $(I-K)\sim 2$--$4$.

We are using ultra-deep 1.4-GHz VLA maps to confirm the reliability of
the identifications\cite{I99b} and find radio counterparts brighter
than $\sim 50\,\mu$Jy (equivalent to intrinsic flux densities of $\gs
20$\,$\mu$Jy) for over 60\% of the sub-mm sources.  Again this success
rate is similar to that found for sub-mm/radio comparisons in the {\it
HDF}\cite{R98}.  The radio fluxes of the SCUBA sources are broadly
in line with those expected if the sub-mm emission is powered by
starbursts and the sources follow the locally determined $L_{\rm
FIR}$--$L_{\rm 5 GHz}$ correlation for galaxies.  Similar analysis of
the other surveys await deep radio observations and confirmation of the
astrometric accuracy of the SCUBA maps. 
  
%
% Figure 3
%
\begin{figure}[htb]
\centerline{\epsfig{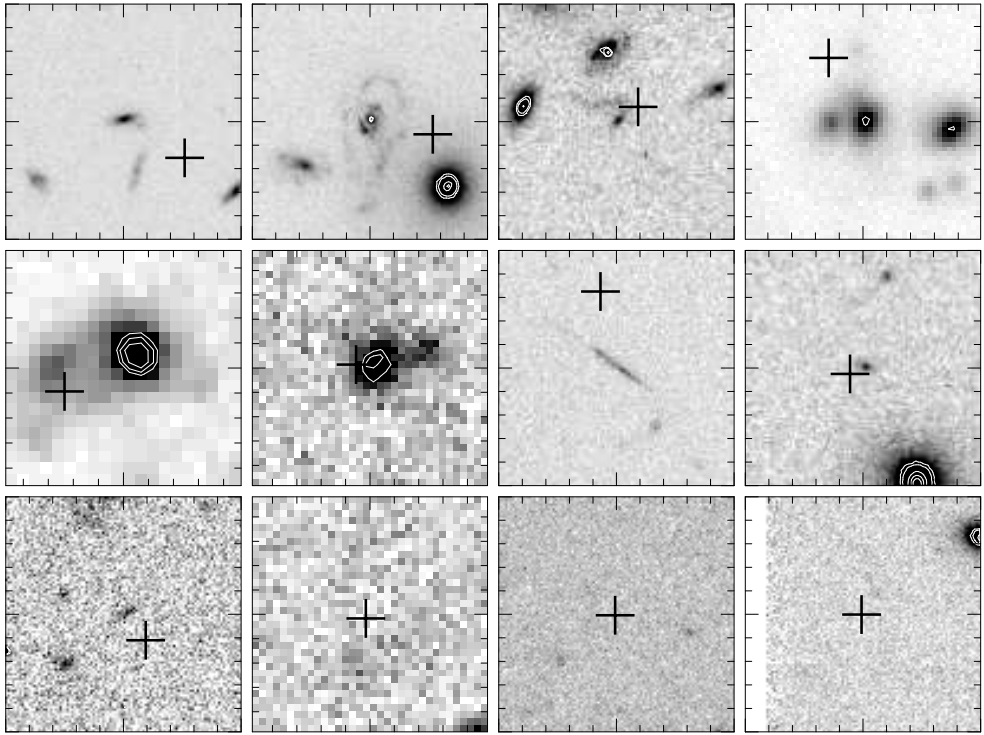}}
\vspace*{0.3cm}
\caption{
The 12 optically-faint sub-mm galaxies from the SCUBA Lens Survey sample
for which we have high-resolution $R$- or $I$-band imaging [12].  Each panel is
$10\times 10$\,arcsec corresponding to $\gs 80$\,kpc at $z>1$ and they
are ordered from the upper-left on the basis of their morphologies: 6
disturbed/interacting, 4 compact/featureless (including a
strongly-distorted arclet) and 2 blank fields.  The images are centred
on the most likely optical candidate where one is known. The centroids
of the sub-mm sources are indicated by crosses.  Note that these images
span a range in exposure times and resolutions [12].
}
\end{figure}

The morphologies of those galaxies for which we have high-resolution
optical imaging fall into three broad categories: faint disturbed
galaxies and interactions; faint galaxies too compact to classify
reliably; and dusty, star-forming galaxies at intermediate
redshifts\cite{SIBK,L98}.  We show in Fig.~3 the $R$- or $I$-band
images of the optically faint sub-mm galaxies, these illustrate the
dominance of disturbed morphologies in this sample.  About 70\% of the
faint galaxies are disturbed or interacting, suggesting that in
the distant Universe, as in the local one, interactions remain an
important mechanism for triggering starbursts and for the formation of
ULIRGs\cite{SIBK}. The  faint, compact galaxies may represent a later
evolutionary stage of these mergers, or more centrally concentrated
starbursts. It is likely that some of these will also host active
galactic nuclei.

\subsection*{The Redshift Distribution of Faint Sub-mm Galaxies}
\vspace*{-2mm}
An analysis of the optical colors of our sub-mm sample to search for
the signature of the Lyman break in bluer passbands allows us to
estimate a crude redshift distribution from their identification in
deep $B$- and $V$-band images\cite{SIBK}.  This indicates that $\gs
75$\% of the optically-identified galaxies have $z\ls 5.5$ whilst $\gs
50$\% lie at $z\ls 4.5$ on the basis of $B$-band identifications
alone.  Photometric redshifts have been used in the {\it
HDF}\cite{Hu98} and CFRS\cite{L98} to place limits on the redshift
distributions of sub-mm galaxies in these samples, the estimated
redshifts span $z\sim 1$--4.  Although the reliability of the
photometric redshifts obtained for such dusty systems has yet to be
tested, these results are consistent with the limits above.  We
conclude that the luminous sub-mm population is broadly coeval with the
more modestly star-forming galaxies selected by UV/optical surveys of
the distant Universe (e.g.\ \cite{Mad}).  However, the individual SCUBA
galaxies have SFRs which are typically an order of magnitude higher
than those of the optically selected galaxies, as well as being
apparently more dust (and hence metal?) rich.  

A further attraction of using lenses in our survey was the possibility
of deriving redshift estimates for galaxies too faint for spectroscopic
identification.  Using our detailed mass models, redshifts can be
determined for any background galaxy whose distortion can be measured from our
{\it HST} imaging, using the relationship between source redshift and
apparent shear for the lens model\cite{JPK96}, a technique whose
accuracy has been recently confirmed\cite{TE98}.  In this way we have
estimated a redshift of $z=1.6\pm 0.2$ for an arclet detected in our
survey with an intrinsic apparent magnitude of $I=25.3$.

Preliminary results from spectroscopic surveys of the different sub-mm
samples are beginning to appear, although none are yet complete.  First
indications from the existing CFRS redshift survey\cite{L98}, for which
the majority of galaxies have $z<0.8$, unsurprisingly finds a low
redshift for the identified sources, $z\sim 0.1$--0.7.  Spectroscopic
observations of the SCUBA Lens Survey sample with Keck, CFHT and WHT
have identified a number of distant galaxies in the range $z\sim
1$--3\cite{B99} as well as several galaxies at $z\sim 0.2$--0.4 in the
foreground cluster lenses\cite{B99,E98}. These sources are removed from
our count analysis, but their detection does confirm that SCUBA has the
ability to routinely detect star-forming galaxies  at $z<1$\cite{L98}.
The more distant systems include a $z=2.8$ dusty
AGN/starburst\cite{I98} and a $z=2.6$ starburst\cite{B99,I99a}.   These
spectroscopic observations support the photometrically-derived redshift
limits for the bulk of the population as well as giving more
information about the dominant emission processes in individual
galaxies. In particular, the spectra provide some indication of the
relative fractions of AGN and starbursts in the sub-mm population.
Nevertheless, the spectroscopic surveys currently have nothing to say
about the 10--20\% of SCUBA sources that have no obvious optical
counterpart.  Owing to the large negative $K$-correction in the sub-mm,
these sources may be at $z>5$, and thus are the most interesting
sources to followup.

%
% Figure 4
%
\begin{figure}[htb]
\centerline{\epsfig{file=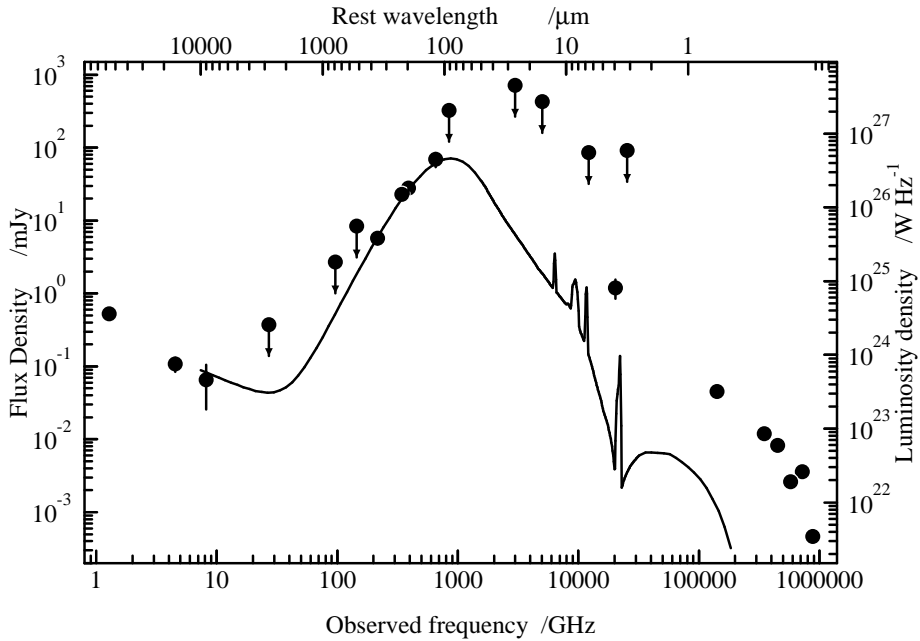,width=11.0cm}}
\vspace*{0.3cm}
\caption{
The spectral energy distribution (SED) of SMM\,J02399$-$0136 at
$z=2.8$, adapted from [26].  The intrinsic FIR luminosity of this
source is 10$^{13}$\,L$_\odot$, after correction for lens amplification
using our lens model, making it one of the most luminous galaxies
known. The strong peak around 100\,$\mu$m in the restframe, probed by
the SCUBA observations at 450\,$\mu$m--2\,mm, is emission from dust at
40--50\,K in the galaxy.  The radio emission from this galaxy is at a
level expected from the high SFR, as shown by the agreement with the
rescaled composite SED of luminous {\it IRAS} galaxies plotted here [27].
}
\end{figure}

\subsection*{The Physical Properties of Sub-mm Galaxies}
\vspace*{-2mm}
Further insights into the nature of the sub-mm-selected galaxies come
from detailed investigations of the physical properties of this
population (e.g.\ SFR, T$_{\rm dust}$, M$_{\rm dust}$, M$_{\rm gas}$,
etc.).  These studies are not easy though, in particular turning the
observed sub-mm flux into a SFR or dust mass requires a number of
uncertain steps, even in the absence of competing contributions from an
AGN and a starburst within a single galaxy.  The basis for all of these
analyses are multi-wavelength observations, especially in the sub-mm.
Fortunately the 450/850-$\mu$m arrays and 1.3- and 2.0-mm
single-channel bolometers on SCUBA can provide the necessary sub-mm
data to constrain quantities such as the dust temperature in
high-redshift galaxies.  Radio observations are useful not only to
provide more accurate positions, but also to rule out non-thermal
contributions to the sub-mm flux.  

A well-sampled SED is shown in Fig.~4 for the brightest source in the
SCUBA Lens Survey, SMM\,J02399$-$0136 at $z=2.8$\cite{I98}.  The
amplification-corrected FIR luminosity of this galaxy is
10$^{13}$\,L$_\odot$, its dust temperature is T$_{\rm dust} =
40$--$50$\,K and the dust mass is around M$_{\rm dust} \sim 6\times
10^8$\,M$_\odot$.  The estimated SFR for this galaxy is $\gs
2000$\,M$_\odot$ yr$^{-1}$, although the optical spectrum indicates
that it hosts a Seyfert-2 nucleus, suggesting that some fraction of the
L$_{\rm FIR}$ is probably attributed to the AGN\cite{I98}.
SMM\,J02399$-$0136 has recently been detected in CO using the Owens
Valley Millimeter Array\cite{F98}, revealing the presence of $\sim 2
\times 10^{11}$\,M$_\odot$ of molecular gas, a dynamically important
component of this massive galaxy and an enormous reservoir of fuel to
power the star-formation.  At this time, SMM\,J02399$-$0136 is the only
sub-mm-selected galaxy for which such detailed observations are
available.

\subsection*{The Nature of the Faint Sub-mm Population}
\vspace*{-2mm}
We are beginning to build up a picture of the population of distant,
luminous sub-mm galaxies which dominate the FIRB at wavelengths around
1\,mm.  The fact that the sub-mm population detected by SCUBA can
account for all of the {\it COBE} background indicates that a
substantial fraction (upto half) of the stars in the local Universe
could be formed in these systems.  The properties of this population
are similar to those of ULIRGs in the local Universe, with the
important distinction that they contribute a sub-mm luminosity density
at early epochs that is more than an order of magnitude greater than
the corresponding galaxies today.  Detailed observations of this
population can be used to trace the amount of high-redshift
star-formation activity that is obscured from view in the optical by
dust, and so is missing from existing inventories of star-formation
activity in the distant Universe\cite{SIB,Hu98}. In this way a more complete
and robust  history of star formation for the Universe can be constructed.  

As with local ULIRGs, there is uncertainty over the exact contributions
from AGN and starbursts to this luminosity density.  Insight may come
through searches for hard X-ray emission from the sub-mm galaxies using
{\it AXAF}, as these should detect all but the most heavily enshrouded
AGN at $z>1$\cite{KFG}.  These searches will also provide an estimate
of the total contribution from the dust-obscured AGN to the X-ray
background\cite{KFG,A98}.  At the current time we can only state that
$\gs 20$\% of the sub-mm population shows obvious spectral signatures
of an AGN --- although this does not mean that the AGN dominates the
emission in the sub-mm.  Assuming that the bulk of the sub-mm emission
arises from starbursts, we find that the total amount of energy emitted
by dusty galaxies is about four times greater than that inferred from
rest-frame UV observations, and that a larger fraction of this energy
is emitted at high redshifts\cite{BSIK}. The simplest explanation for
these results is that a large population of luminous strongly obscured
galaxies at redshifts of $z\ls 5$ is missing from optical surveys of
the distant Universe.

Finally, we come to the question of what class of object the sub-mm
galaxies will evolve into by the present day?  The similarities of
these sources to local ULIRGs, which are expected to evolve into
elliptical galaxies\cite{M96}, along with their apparently high SFRs,
which if sustained over $\sim 1$\,Gyr would form an entire $L^\ast$
galaxy at high redshift, suggest that the sub-mm population may be
young, massive elliptical galaxies\cite{SIB,Ea98}.  A question then
arises concerning the relationship between the sub-mm population and
the other class of high-redshift source which has been identified with
forming ellipticals: the Lyman-break galaxies\cite{St96}.  These
objects appear to have typically lower SFRs and less dust than the
sub-mm galaxies.  The two populations could be naturally linked if the
dust content of young galaxies is coupled to their masses or
luminosities\cite{Di98}, such that the more massive galaxies are
dustier. This behaviour would parallel the metallicity relation for
elliptical galaxies, which show increasing metal enrichment with
increasing  galaxy mass, arising from the  retention of processed gas
by the potentials of the more massive galaxies. We would therefore
identify the most massive, young ellipticals with the sub-mm galaxies
and the less massive ellipticals and bulges with the Lyman-break
objects.  Thus detailed observations of the sub-mm population should
give much needed observational information for models of the formation
and evolution of massive galaxies\cite{BSIK}.  In particular we look
forward to further CO detections of sub-mm galaxies with both the
existing and next-generation millimeter arrays to study the kinematics
of these systems and hence determine their masses.
  
\subsection*{Acknowledgements}
\vspace*{-2mm}
We thank Amy Barger and Len Cowie for the use of results from our
on-going collaborative program of spectroscopy with the Keck telescopes
and Omar Almaini, Steve Eales, Katherine Gunn, Dave Hughes, Andy
Lawrence, Malcolm Longair, Chris Mihos, Ian Robson and Michael
Rowan-Robinson for useful conversations.  IRS thanks the organisers for
support while at the conference and acknowledges a Royal Society
Fellowship.
\vfil\eject

\end{document}